\documentclass[conference]{IEEEtran}
\usepackage{amsmath,amssymb,amsfonts}
\usepackage{graphicx}
\usepackage{textcomp}
\usepackage{lineno}
\usepackage{soul}
\sethlcolor{BurntOrange}

\newcommand{\rdel}[1]{}

\input{preamble.tex}

\IEEEoverridecommandlockouts

\begin{document}

\title{Nonlinear spectral clustering with C++ GraphBLAS
}

\author{Dimosthenis Pasadakis \& Olaf Schenk \& Verner Vlacic \& Albert-Jan Yzelman
\thanks{Dimosthenis Pasadakis, and Olaf Schenk are with the Advanced
Computing Laboratory at the Institute of Computing, Universit\`a della
Svizzera italiana (USI), Lugano, Switzerland. email: 
 \{dimosthenis.pasadakis, olaf.schenk\}@usi.ch. 
 Verner Vlacic, and Albert-Jan Yzelman are with the Computing Systems Lab, Huawei Zurich Research Center, Switzerland. email: 
 \{verner.vlacic, albertjan.yzelman\}@huawei.com.}%

}

\maketitle

\begin{abstract}
Nonlinear reformulations of the spectral clustering method
have gained a lot of recent
attention due to their increased numerical benefits and
their solid mathematical background. However,
the estimation of the multiple nonlinear eigenvectors is
associated with an increased computational cost.
We present an implementation of a direct multiway spectral
clustering algorithm in
the $p$-norm, for $p\in(1,2]$, using a novel C++ GraphBLAS API. The 
key operations are expressed in linear algebraic terms and
are executed over the resulting sparse matrices and dense
vectors, parameterized in the algebra pertinent to the
computation.
We demonstrate the effectiveness and accuracy of our shared-memory algorithm on several artificial test cases. Our numerical examples and comparative results 
against competitive methods indicate that the proposed
implementation attains high quality clusters in terms of the balanced
graph cut metric. The strong scaling capabilities of our algorithm are showcased
on a range of datasets with up to $8$ million nodes and $48$ million edges.
\end{abstract}


\section{Introduction}
\label{ref:intro}

Spectral clustering is a popular community
detection method that can be applied
to any kind of data
with a suitable similarity metric between them forming a graphical
structure. At its core lies the computation of the mutually orthogonal
eigenvectors of the graph Laplacian, a symmetric and positive
semi-definite matrix, which
are treated as the spectral coordinates of the graph, and are
subsequently
discretized using distance based algorithms~\cite{Luxburg07}. 
This eigenspectrum computation offers ample room for
parallelization, with both shared and distributed memory
implementations widely used~\cite{Wierzchon18}.
Nonlinear variants of the
method in the $p$-norm, for $p \in (1,2]$, that have
been proposed
lead to a
minimization of
balanced graph cut metrics, and an increase in the
accuracy of the final clustering
assignment~\cite{Luo2010}.
Recently
in~\cite{Pasadakis22}, $p$-spectral clustering was cast as a
nonlinear
unconstrained optimization problem on the Grassmann
manifold~\cite{Edelman99}, by
approximating the constraint for $p$-orthogonality with an analogous one for
$2$-orthogonality. This approach is not applicable to large-scale data, due to the
large number of
multiplications of the graph adjacency matrix with the computed
eigenvectors that are required for convergence, especially as the value of $p$ tends to 1.

GraphBLAS is a standard~\cite{Kepner15} for expressing
graph computations in the language of linear algebra. Its core
concepts are (i) algebraic containers, which correspond to sparse
matrices and vectors, (ii) algebraic operators, describing sparse
matrix-vector (SpMV) multiplications, and (iii)  algebraic
relations, an example of which is a generalized semiring under
which an SpMV multiplication takes place. The recently introduced
C++11 implementation of GraphBLAS~\cite{yzelman20} has showcased impressive results
on the speed-up of algorithms based on SpMVs~\cite{scolari23a}.
We express the key operations of the method introduced in~\cite{Pasadakis22} as well as the k-means discretization of the resulting eigenvectors in this C++ GraphBLAS API. This allows us to leverage its auto-parallelisation capabilities, and furnish the first,
according to our best knowledge, $p$-norm spectral clustering algorithm applicable to large-scale data for shared-memory machines.


\section{A C++ GraphBLAS $p$-spectral clustering algorithm}
\label{sec:pGrass_method}

For an undirected weighted graph $\mathcal{G}(V,E,\bb{W})$ where $V$
is the set of $n$ nodes, $E$ the set of edges, and $\bb{W}$ the
weighted adjacency matrix, estimating a set of $k$
$p$-eigenvectors on the Grassmann manifold $\mathcal{Gr}$
can be expressed as
\begin{equation}
\label{eq:Grassman_Opt_Problem}
    \minimize{\bb{U} \in \mathcal{Gr}(k,n)}\;   F_p(\bb{U}) = \sum_{\ell=1}^k \sum_{i,j = 1}^n \frac{ w_{ij} |u_i^\ell - u_j^\ell|^p}{2\|\bb{u}^\ell\|^p_p}, \; \; p \in (1,2].      
\end{equation}
Let 
$\ell = 1, 2, \dots, k$ denote the eigenvector indices, and,  at the minimizer, 
the columns of $\vect{U} = \left( \vect{u}_1, \dots, \vect{u}_k
\right)$ approximate the eigenvectors associated with the smallest
$k$ eigenvalues of the $p$-Laplacian operator $\bb{\Delta}_p$. For $i \in V$ the $p$-Laplacian operator is defined as
$
    \left(\bb{\Delta}_p \vect{u} \right)_i = \sum_{j \in V} w_{ij} \phi_p \left( u_i - u_j \right),
$
with $\phi_p : \mathbb{R} \rightarrow \mathbb{R}$  being
$
    \phi_p(x) = |x|^{p-1} \text{sign}(x),
$
and the $p$-norm is $\| \bb{u} \|_p = \sqrt[p]{ 
\sum_{i=1}^n |u_i|^p}$, for $u \in \mathbb{R}$.

We use the Riemannian optimization software package 
ROPTLIB~\cite{Huang18} to minimize~\eqref{eq:Grassman_Opt_Problem} for progressively smaller values of $p$ using Newton's method on the Grassmann manifold, where the solution of the linearized Newton subproblems is handled by a truncated conjugate gradient scheme. The description of the optimization problem is accomplished in ROPTLIB by specifying the function \texttt{EucGrad} which computes the gradient of $F_p(\bb{U})$ as well as the function \texttt{EucHessianEta} which computes $\bb{\eta}\mapsto{\mathcal{H}}\bb{\eta}=(\mathcal{H}^\ell\eta^\ell)_{\ell = 1}^k$ for arbitrary $\bb{\eta}\in \mathbb{R}^{k\times n}$, where the collection of matrices $\mathcal{H}^1,\dots, \mathcal{H}^k\in \mathbb{R}^{n\times n}$ corresponds to the Hessian of $F_p(\bb{U})$. For illustration, we include our C++ GraphBLAS implementation of \texttt{EucHessianEta} in Algorithm~\ref{alg:euchessianeta}.
Here the subroutines ROPTLIBtoGRB and GRBtoROPTLIB serve for I/O from and to the ROPTLIB data structures. The C++ GraphBLAS API leverages the algebraic structure of the ring of real numbers to parallelize the SpMV operation (the grb::vxm primitive).

\begin{algorithm}
        \flushleft
        \begin{tabular}{rl}
                Input: &
                \begin{tabular}{r@{ }l}
                        &$\bb{\eta}$, a $k\times n$ matrix\\
                        &$(D[\ell])_{\ell=1}^k$, where each $D[\ell]=\mathrm{diag}(\mathcal{H}^\ell)$  \\
                        &$(H[\ell])_{\ell=1}^k$, where each $H[\ell]= \mathrm{diag}(\mathcal{H}^\ell)-\mathcal{H}^\ell $
                \end{tabular} \\
                Output: & $\bb{r}$, the result of $\bb{\eta}\mapsto{\mathcal{H}}\bb{\eta}$\\
        \end{tabular}

        \begin{algorithmic}[1]
            \STATE grb::Semiring$<$grb::operators::add$<$double$>$, \\\hspace{1em}grb::operators::mul$<$double$>$,\\ \hspace{1em}grb::identities::zero, grb::identities::one$>$ reals\_ring;
            \STATE std::vector$<$grb::Vector$<$double$> >$ grb\_eta, grb\_res;
            \STATE grb::Vector$<$double$>$ v, w;
            \STATE ROPTLIBtoGRB($\bb{\eta}$, grb\_eta);
            \FOR{$\ell = 1$\textbf{ to }$k$}
            \STATE grb::set(v, 0); 
            \STATE grb::vxm(v, grb\_eta[$\ell$], H[$\ell$], reals\_ring);
            \STATE grb::eWiseApply(w, grb\_eta[$\ell$]),D[$\ell$],\\ \hspace{1em}grb::operators::mul$<$double$>$());
            \STATE grb::eWiseApply(grb\_res[$\ell$], w, v,\\ \hspace{1em}grb::operators::subtract$<$double$>$());
            \ENDFOR
            \STATE GRBtoROPTLIB(grb\_res, $\bb{r}$);
            \RETURN $\bb{r}$
        \end{algorithmic}
        \caption{The function \texttt{EucHessianEta}.}
        \label{alg:euchessianeta}
\end{algorithm}
\section{Numerical Results}
\label{sec:num_res}

In order to demonstrate the effectiveness of the C++ GraphBLAS API for implementing the
$p$-spectral clustering method, in Section~\ref{sec:pGrass_cut} we report on the quality of
the graph cuts obtained, and in Section~\ref{sec:pGrass_par_perf} we present its parallel performance.
For our experiments we select $8$ matrices from the 
SuiteSparse
matrix collection~\cite{Davis11} with an increasing number of nodes $n =
2^r$ and edges $m \approx 6 * 2^r$, for $r = 16, \dots , 23$,
corresponding to Delaunay triangulations in a
unit square.

\subsection{Quality of graph cuts}
\label{sec:pGrass_cut}

We identify four clusters $C_i$, $i=1,2,3, 4$, and compute the value of the balanced
graph cut metric
$\mathrm{RCut}(C_1, C_2, C_3, C_4) = \sum_{i = 1}^4 \frac{\mathrm{cut}(C_i,\overline{C_i})}{|C_i|}$. The
results for the mid-scale cases with
$r = \{16, 17, 18, 19\}$ are summarized in Table~\ref{tab:RCut_improve}. We compare our method
(GrB-$p$Grass) against traditional spectral clustering (Spec)~\cite{Luxburg07} and against the
first full eigenvector analysis of $p$-Laplacian leading to direct multiway clustering ($p$Multi)~\cite{Luo2010}.

\begin{table}[!t]
    \centering
    \resizebox{\linewidth}{!}{%
\begin{tabular}{l|c|c|c|c}
    \toprule
    Method          &   Del. 16           &    Del. 17           &  Del. 18         &   Del. 19          \\
    \midrule
    \midrule
    Spec            & $0.129$         &  $0.089$        &   $0.062$        &  $0.045$       \\
    $p$Multi        & $-6.21\%$       &  $-3.11\%$      &   $-2.21\%$      &  $-2.45\%$     \\
    GrB-$p$Grass    & $\bb{-8.12}\%$  &  $\bb{-6.43}\%$ &   $\bb{-4.56}\%$ &  $\bb{-4.19}\%$ \\
    \bottomrule
\end{tabular}
}
\caption{\label{tab:RCut_improve} 
Results in terms of the balanced graph cut metric RCut. We report the baseline (Spec) RCut, and
in percentage the reduction of the cut that the methods pMulti and GrB-$p$Grass (ours) achieved.
}
\end{table}

\subsection{Parallel performance}
\label{sec:pGrass_par_perf}

\begin{figure}[t]
    \centering
\begin{subfigure}[b]{0.49\linewidth}
   \includegraphics[width=.99\columnwidth]{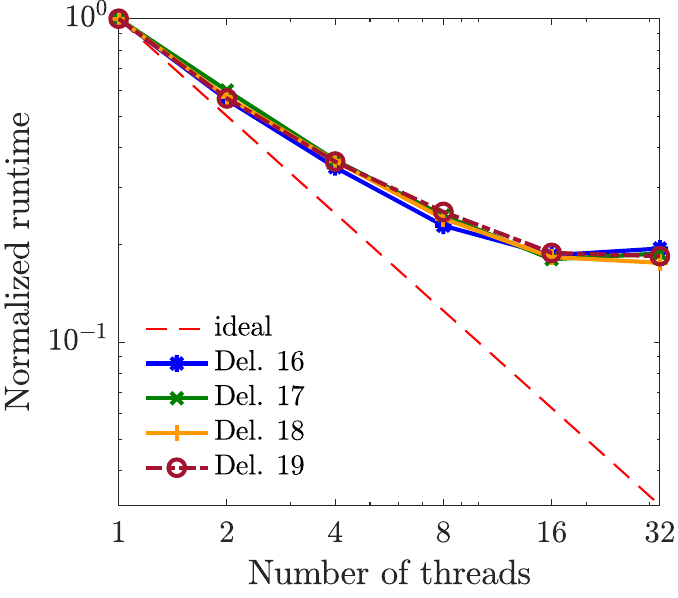}
   \caption{}
   \label{fig:Scale_med} 
\end{subfigure}
\begin{subfigure}[b]{0.49\linewidth}
   \includegraphics[width=.98\columnwidth]{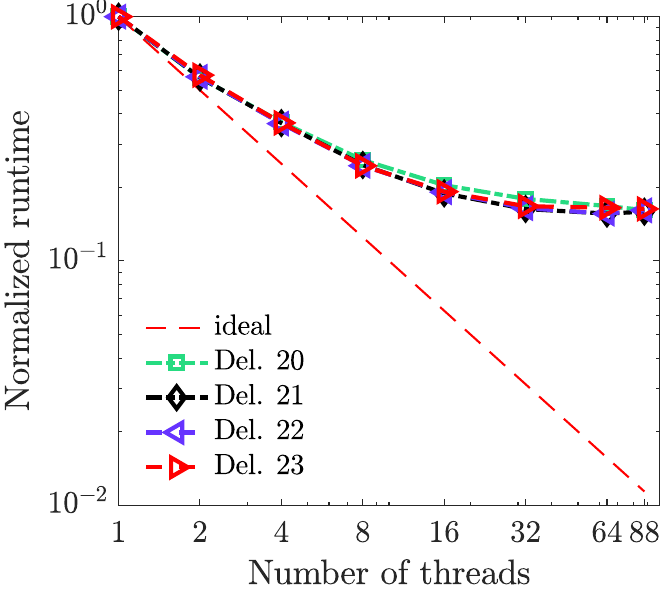}
   \caption{}
   \label{fig:Scale_large}
\end{subfigure}
    \caption{Strong scaling of the C++ GraphBLAS components of the algorithm for the Delaunay graphs.
    a) Results for the mid-scale cases of node size $n \in [2^{16}, 2^{19}]$,
    b) Results for the large-scale cases of node size $n \in [2^{20}, 2^{23}]$.
    The runtime is normalized versus single-thread execution.
    } 
    \label{fig:Strong_Scale_Grb}
\end{figure}

The strong scaling results of the developed algorithm are 
illustrated in Figure~\ref{fig:Strong_Scale_Grb}. In both plots,
the dashed red line indicates the ideal scalability. We utilize up to
$32$ threads for the mid-scale cases  (Figure~\ref{fig:Scale_med}),
and up to $88$ threads for the large-scale Delaunay graphs with $r = \{20,21,22,23\}$
(Figure~\ref{fig:Scale_large}). On average, the parallel execution of the
algorithm is $5.5\times$ faster than its sequential variant for the 
mid-scale tests, and $6.4\times$ faster for the large-scale cases. The run-time of the smallest
case ($r = 16$) was $\thicksim300$ sec, and that of the largest
one ($r = 23$) $\thicksim 20$ hours. A breakdown of the runtime
shows that only the GraphBLAS components of the algorithm exhibit excellent
weak scalability for the large-scale graphs.

\section{Conclusion \& Outlook}
\label{ref:conclusion}

In this work, we have expressed they key operations of a multiway $p$-spectral clustering  algorithm in the
C++ GraphBLAS API. This enabled accurate parallel clustering of large-scale graphs on a
shared-memory machine. We intend to further explore the potential gains of expressing graph
partitioning and clustering algorithms in linear algebraic terms.

\section*{Acknowledgment}
D.P. and O.S. acknowledge the support of the joint 
DFG - 470857344 and SNSF - 204817 project.
\bibliographystyle{IEEEtran}
\bibliography{pLap_GrB_ref}

\end{document}